\documentclass[12pt]{iopart}
\usepackage{graphicx}
\usepackage{amssymb}
\usepackage{bm}

\begin{document}
\title{Resonant, non-resonant, and anomalous  states  of Dirac electrons  in a  parabolic well in the presence of magnetic fields}
\author{S. C. Kim, J. W. Lee, and S. -R. Eric Yang$^{*}$}
\address{ Physics Department, Korea  University, Seoul Korea 136-713}
\ead{$^{*}$corresponding author eyang812@gmail.com}

\begin{abstract}
We report on  several new basic properties of  a  parabolic dot in
the presence of a magnetic field. The ratio between  the potential
strength and the Landau level (LL) energy spacing serves as the
coupling constant of this problem. In the weak coupling limit the
energy spectrum in each Hilbert subspace of an angular momentum
consists of discrete LLs of graphene. In the intermediate coupling
regime  non-resonant states form a closely spaced energy spectrum.
We find, counter-intuitively, that resonant quasi-boundstates of
both positive and negative energies exist in the spectrum. The
presence of resonant quasi-boundstates of negative energies is a
unique property of massless Dirac fermions. As the strong coupling
limit is approached resonant and non-resonant states transform into
anomalous states, whose probability densities develop a narrow  peak
inside the well and another broad peak under the potential barrier.
These properties may investigated experimentally by measuring
optical transition energies that can be described by a scaling
function  of the coupling constant.
\end{abstract}

\section{Introduction}

Two-dimensional parabolic quantum dots of semiconductor
heterostructures have been studied widely\cite{Heit, McE} both experimentally and
theoretically because they are excellent candidates for single
electron transistors. They can effectively confine electrons   and
the number of electrons in them can be controlled using a gate
potential.
Their Hamiltonian is
\begin{eqnarray}
H= \frac{1}{2m}({\vec p}+\frac{e}{c}{\vec A}
)^2+\frac{1}{2}m\Omega^2 r^2
\end{eqnarray}
with a magnetic field $\vec{B}$ is applied perpendicular to the 2D plane (vector potential $\vec A$ is given
in a symmetric gauge). The characteristic length scale of the problem is given by
$\lambda^2=\frac{\hbar}{m\sqrt{4\Omega^2+\omega_c^2}}$, where
$\omega_c=\frac{eB}{mc}$ is the cyclotron frequency.  This problem can be solved exactly\cite{Fock}, and the
eigenenergies are all  positive and their  spectrum is discrete.

Massless Dirac electrons\cite{Mac0} moving in a 2D parabolic potential display several different
features in comparison to massful electrons.
They are described by the Dirac Hamiltonian
\begin{eqnarray}
H= v_F\vec{\sigma}\cdot ({\vec p}+\frac{e}{c}{\vec A}
)+\frac{1}{2}\kappa r^2.
\end{eqnarray}
 No exact solutions of this problem are known
in graphene and several fundamental properties are still unknown,
such as the existence of resonant and non-resonant states. These
basic properties  may affect the experimentally relevant optical
spectrum in a profound way. The dimensionless  coupling constant of
this problem is the ratio between the strength of the potential
$\frac{1}{2}\kappa\ell^2$ and the LL energy separation
$E_C=\frac{\hbar v_F}{\ell}$
\begin{eqnarray}
\alpha=\kappa\ell^2/E_C=\frac{\kappa\ell^3}{\hbar v_F},
\end{eqnarray}
where the magnetic length is $\ell=\sqrt{\hbar c/Be}$.  One is in
the strong coupling regime $\alpha\gg 1$ for small value of $B$ or
large value of potential strength $\kappa$. Parabolic  dots in
magnetic fields have been investigated numerically in the weak
coupling regime $\alpha<1$. The energy spectrum is found to be
discrete, and eigenstates are quasi-boundstates with long
oscillating tails under the barrier\cite{gia,Chen,Park3}.  Also some
of these states   exhibit anticrossings\cite{gia,Park3}.  The
parabolic potential acts as a singular perturbation\cite{Ben}
because eigenstate wavefunctions are qualitatively different from
those in the absence of a parabolic potential.

The problem has not been investigated away from the weak coupling
regime.  It is a highly  non-trivial problem. This can be seen as
follows. One of the special features of graphene LLs is the presence
of negative energy states under the potential barrier\cite{Park3}.
The first order energy correction of a LL state $\psi_{n,m}(r)$ is,
for sufficiently large $n$,
\begin{eqnarray}
\nonumber\\
\langle\psi_{n,m}|V(r)|\psi_{n,m}\rangle\sim\kappa\langle r^2
\rangle\sim\kappa\ell^2 |n|. \label{eq:selfener}
\end{eqnarray}
This result suggests that  a LL state with a large {\it negative}
energy, $-E_C\sqrt{2|n|}$ with $|n|\gg 1$, corresponding to having a
large average radius $\sqrt{\langle r^2 \rangle}$, acquires a
significant positive energy correction, which can make the
renormalized energy {\it positive}. In the dimensionless units this
energy correction is $\kappa\ell^2|n|/E_C=\alpha |n|$, which
suggests that even for small value of $\alpha$ the correction can be
significant for $|n|\gg 1$. Moreover, it is unclear how eigenstates
evolve from weak to strong coupling regimes. A simple dimensional
analysis suggests that the energy scale of the problem in the strong
coupling limit of $B\rightarrow 0$ or $\alpha\rightarrow \infty$ is
$\kappa^{1/3}(\hbar v_F)^{2/3}$.  In units of $E_C$ this energy
scale is $\alpha^{1/3}$. It indicates that the dimensionless energy
level spacing increases from $\sim 1$ to $\sim\alpha^{1/3}$ as one
moves from weak to strong coupling regimes. However, studies in
ordinary semiconductors suggest that the crossover regime may be
non-trivial \cite{Mac}.

\begin{figure}[!hbpt]
\begin{center}
\includegraphics[width=0.35\textwidth]{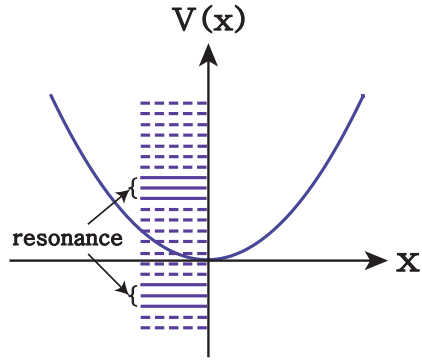}
\caption{ Schematic energy spectum  of a parabolic dot in the
intermediate coupling regime $\alpha\sim 1$.  Energy spectrum
 is closely spaced, and resonant states of
positive and negative energies are present.}\label{fig:sch_enspec}
\end{center}
\end{figure}

We have investigated these issues by solving large Hamiltonian
matrices.  Let us give a brief summary of our results in a Hilbert
subspace of   angular momentum $J$. We have studied how eigenvalues
and eigenstates evolve as $\alpha$ increase and find that they
change in a non-trivial way. In the weak coupling limit  of
$\alpha\rightarrow 0$ the spectrum consists of discrete LLs. In the
intermediate coupling regime $\alpha\sim 1$ non-resonant states form
a closely spaced  energy spectrum, see Fig.\ref{fig:sch_enspec}. We
find, counter-intuitively, that resonant quasi-boundstates of both
positive and negative energies exist. The presence of resonant
quasi-boundstates of negative energies is a unique property of
massless Dirac fermions, but they are well-defined  only for $\alpha
< 1$. In the strong coupling regime $\alpha\gg 1$ both resonant and
non-resonant states transform into {\it anomalous} states, and a
sharp distinction between resonant and non-resonant states  no
longer exists. Probability densities of anomalous states develop  a
narrow peak inside the well and decays slowly with small
oscillations under the barrier. The energy level spacing between
them is proportional to the value $\kappa^{1/3}(\hbar v_F)^{2/3}$
and is independent of $\ell$. We show that optical transition
energies between resonant quasi-boundstates can be described by  a
scaling function  of $\alpha$.

\section{ Basis states and Hamiltonian matrix}

In our Hamiltonian matrix
approach the basis states are chosen as graphene LL states
$\psi_{n,m}(\vec{r})$ with  two components A and B
\begin{eqnarray}
\psi_{n,m}(\vec{r})=c_{n}\left(\begin{array}{c}-\textrm{sgn}(n)i\phi_{|n|-1,m}(\vec{r})\\
\phi_{|n|,m}(\vec{r})\end{array}\right)\label{twocomp}.\label{eq:spinor}
\end{eqnarray}
Here $\textrm{sgn}(n)=-1,0,1$ for $n<0, n=0, n>0$ with $n$ and $m$
integers ($m\geq 0$), and $c_n=1$ for $n=0$ and $1/\sqrt{2}$
otherwise.
These basis states can have
{\it positive} or {\it negative} LL energies:
\begin{eqnarray}
E_n=\textrm{sgn}(n)E_C\sqrt{2|n|}.
\end{eqnarray}
The wavefunctions $\phi_{n,m}(\vec{r})$ are the Landau
level wavefunctions of ordinary two-dimensional systems\cite{Yosi}
\begin{eqnarray}
\phi_{n,m}(\vec{r})&=&A_{n,m}\exp\left(i(n-m)\theta-\frac{r^2}{4\ell^2}\right)\left(\frac{r}{\ell}\right)^{|m-n|}\nonumber\\
&\times&L_{(n+m-|m-n|)/2}^{|m-n|}\left(\frac{r^2}{2\ell^2}\right),
\end{eqnarray}
where $A_{n,m}$ are the normalization constants and $L_n^m(x)$ are
Laguerre polynomials. In the presence of a parabolic potential
$J=|n|-m-\frac{1}{2}$ remains a good quantum number. The average
radius of $\phi_{n,m}(\vec{r})$ is given by
\begin{eqnarray}
\langle r^2\rangle=2\ell^{2}(n+m+1)=2\ell^{2}(n+|n|-J+1/2).
\end{eqnarray}

\begin{figure}[!hbpt]
\begin{center}
\includegraphics[width=0.4\textwidth]{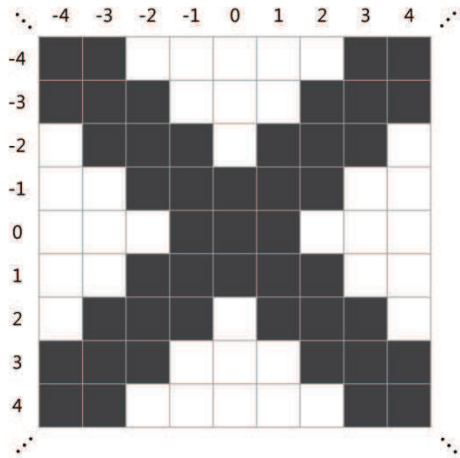}
\caption{Structure of the Hamiltonian matrix  is shown for $J<0$.
For a given $J$ we choose to use $n$ as the basis index instead of
$m$. It  runs from $-(N_c-1)/2$ to $(N_c-1)/2$. Filled (empty)
squares represent non-zero (zero) elements.  For $J>0$ allowed basis
states are given in TABLE II of Ref.\cite{Park2}.} \label{matstruc}
\end{center}
\end{figure}

To investigate the strong coupling effects a large number of basis
states $\psi_{n,m}(r)$ is required.  It is convenient to divide the
Hilbert space into subspaces of angular momentum $J=\pm \frac{1}{2},
\pm \frac{3}{2}, \pm \frac{5}{2},\cdots$.    We diagonalize the
Hamiltonian matrix in each Hilbert subspace $J$. For given $J$, the
matrix elements of the parabolic potential can be written as sum of
two components:
\begin{eqnarray}
&&\langle\psi_{n,m}|\frac{V(r)}{E_C}|\psi_{n',m'}\rangle\nonumber\\
&&=c_n c_{n'}\textrm {sgn}(nn')
\langle\phi_{|n|-1,m}|\frac{V(r)}{E_C}|\phi_{|n'|-1,m'}\rangle
\nonumber\\
&&\ \ + c_n c_{n'}
\langle\phi_{|n|,m}|\frac{V(r)}{E_C}|\phi_{|n'|,m'}\rangle.
\end{eqnarray}
Using the following property of Laguerre polynomials
\begin{eqnarray}
L_n^{\alpha}(x)=\frac{1}{x}\left[(n+\alpha+1)L_n^{\alpha-1}(x)-(n+1)L_{n+1}^{\alpha-1}(x)\right],
\end{eqnarray}
and the orthogonality
\begin{eqnarray}
\int_0^{\infty} x^{\alpha} \textrm{e}^{-x} L_n^{\alpha}(x)
L_m^{\alpha}(x)\, \textrm{d}x
=\frac{\Gamma(n+\alpha+1)}{n!}\delta_{n,m},
\end{eqnarray}
we evaluate the matrix elements.  The resulting matrix is a {\it
sparse} matrix, see Fig.\ref{matstruc}. The dimension of the
Hamiltonian matrix is denoted by $N_c$. When the value of $N_c$ is
sufficiently large the states investigated in this paper do not
exhibit dependence on $N_c$.

\section{Eigenstates of a Hilbert subspace}

Eigenstates of a Hilbert subspace are obtained by diagonalizing the
Hamiltonian matrix. They may be written as a linear combination of
LL wavefunctions with same angular momentum:
\begin{eqnarray}
\Psi_N^J(r)=\sum_n C_n\psi_{n,m}(r). \label{eq:expan}
\end{eqnarray}
Here quantum number $N$ is chosen to be the value $n$ for which
$|C_n|$ is {\it maximum}\cite{com1}.

The following exact results\cite{Park2} are useful in checking
numerical results. The value of wavefunctions at $r=0$ is non-zero
only for $J=-\frac{1}{2}$ and $\frac{1}{2}$:
\begin{eqnarray}
\Psi_N^{J}(0)=\left\{\begin{array}{cc}0 & \textrm{for} \ \ \ J\neq \pm 1/2\\
\textrm{finite} & \textrm{for} \ \ \  J= \pm 1/2.\end{array}\right.
\label{wzero}
\end{eqnarray}
The B and A  components of $|\Psi_{N}^{-1/2}\rangle$  and
$|\Psi_{N}^{1/2}\rangle$  are non-zero  and are of s-wave type.
Their values at $r=0$ can be written as
\begin{eqnarray}
\Psi_{N,B}^{-1/2}(0)&=&\sum_{n\neq 0}
A_{|n|,|n|}C_n/\sqrt{2}+C_0 A_{0,0}\nonumber\\
\Psi_{N,A}^{
1/2}(0)&=&-i\sum_{n}\textrm{sgn}(n)A_{|n|-1,|n|-1}C_n/\sqrt{2}.
\label{mhalf}
\end{eqnarray}
We will only concentrate on optical transitions involving states
with $J=1/2$, $-1/2$, or $-3/2$ since they give the strongest
optical strengths. Unless stated otherwise the results reported in
this paper are for the potential strength  $\kappa =0.1$meV/nm$^2$.

\subsection{ Resonant quasi-boundstates}

\begin{figure}[!hbpt]
\begin{center}
\includegraphics[width=0.35\textwidth]{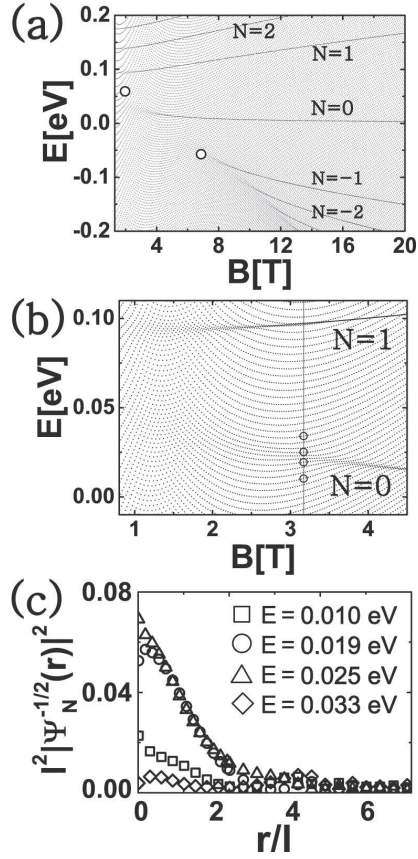}
\caption{ (a) Eigenenergy spectrum of Hilbert subspace of
$J=-\frac{1}{2}$. Lines labeled by $N$ represent resonant
quasi-boundstates.  $2001\times2001$ matrix is used. (b) Enlarged
energy spectrum for $J=-\frac{1}{2}$. Resonant quasi-boundstate
$|\Psi_0^{-1/2}\rangle$ anticrosses strongly  at $B=3.14$T
($\alpha=0.47$). Four circles represent these coupled states. (c)
Probability densities of these four states are displayed. They form
together  a {\it resonance} with the approximate resonant energy
$\epsilon_0^{-1/2}(3.14)=0.025$eV. }\label{fig:Energy_spec}
\end{center}
\end{figure}

\begin{figure}[!hbpt]
\begin{center}
\includegraphics[width=0.3\textwidth]{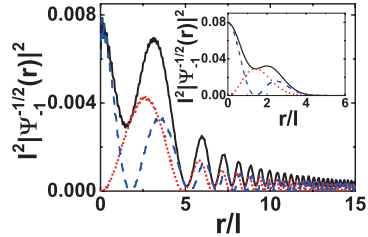}
\caption{ An example of  a resonant quasi-boundstate of negative
energy (indicated as a open circle in Fig.\ref{fig:Energy_spec}(a)).
Dotted (dashed) line represents A (B) components of the probability
density of $|\Psi_{N}^{J}\rangle=|\Psi_{-1}^{-1/2}\rangle$ with
energy $E=-0.060$eV computed at  the coupling constant $B=6.88$T
($\alpha=0.145$). Solid line is the total probability density.
$2001\times2001$ matrix is used. In the absence of the potential the
wavefunction has a peak near $r=0$ (see inset), and its energy is
negative $E=-\sqrt{2}E_C=-0.093$eV. }\label{fig:QuasiBound}
\end{center}
\end{figure}

The computed  energy spectrum of $J=-1/2$ is shown in
Fig.\ref{fig:Energy_spec}(a) for the  range  $0.029<\alpha<0.47$
($3.14T<B<20T$). Spectra for other values of $J=1/2$ and $-3/2$ are
similar, as shown in Sec.IV. Lines labeled by $N$ represent the
energies of resonant quasi-boundstates and other lines represent
non-resonant states. As shown in Fig.\ref{fig:Energy_spec}(b), in
the intermediate coupling regime $\alpha\sim 1$, the energy levels
are closely spaced due to negative energy LL states whose energies
get strongly perturbed upward by the parabolic potential, see
Eq.(\ref{eq:selfener}). A resonant quasi-boundstate anticrosses
other states and becomes strongly mixed with the adjacent states,
see Fig.\ref{fig:Energy_spec}(b).  For example, the resonant
quasi-boundstate $|\Psi_0^{-1/2}\rangle$ is strongly
mixed\cite{com1} at $B=3.14$T ($\alpha=0.47$), and, as shown in
Fig.\ref{fig:Energy_spec}(b), there are three states that could be
identified as $|\Psi_0^{-1/2}\rangle$.  In these states with the
energies $E=0.025, 0.019$, and $0.010$eV the expansion coefficients
$C_n$ of Eq.(\ref{eq:expan}) take the maximum value at $n=0$  with
the values $C_0=0.622, 0.595$, and $0.287$, respectively (Since
$C_0$ is largest for $E=0.025$eV this state is labeled as
$|\Psi_0^{-1/2}\rangle$). Fig.\ref{fig:Energy_spec}(c) displays
probability densities of these states. They form  together a {\it
resonance} with the approximate resonant energy $E=0.025$eV. In the
weak coupling regime the width of a resonance is small and, in order
to resolve it, the energy level spacing must be smaller than the
width of a resonance, which requires a large matrix dimension.

Note also that resonant quasi-boundstates of negative energies
exist. This is a unique property of massless Dirac fermions.  An
example is shown in Fig.\ref{fig:QuasiBound}. They are well-defined
only for sufficiently large $B$, i.e., only in the weak coupling
regime.  The appearance of a second peak away from $r=0$ in the
probability density is different from the usual behavior of the
wavefunction a resonant state.

We see that as $\alpha$ increases, or, as $B$ decreases,  resonant
quasi-boundstates disappear into the closely spaced energy spectrum.
For $N=0$ states this happens around $B\sim 3 T$. For larger values
of $N$ this happens at smaller  values of $B$. We will show in
Sec.III (C) that, as $\alpha$ increases, the peak at $r=0$ increases
and the state becomes anomalous.

\subsection{Non-resonant states}

\begin{figure}[!hbpt]
\begin{center}
\includegraphics[width=0.3\textwidth]{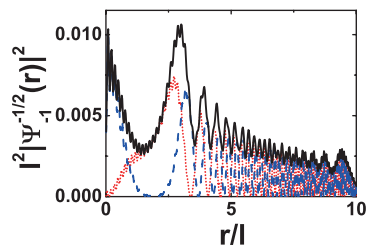}
\caption{An example of  a non-resonant state (indicated as a open
circle in Fig.\ref{fig:Energy_spec}(a)).  Dotted (dashed) line
represents A (B) components of the probability density
$|\Psi_{N}^{J}\rangle=|\Psi_{-1}^{-1/2}\rangle$ with energy $
E=0.054$eV and $B=2.04$T ($\alpha=0.9$). As $r\rightarrow 0$ the
dashed line approaches a finite value while dotted line goes to
zero. $2001\times 2001$ matrix is used. }\label{fig:Anom}
\end{center}
\end{figure}

\begin{figure}[!hbpt]
\begin{center}
\includegraphics[width=0.35\textwidth]{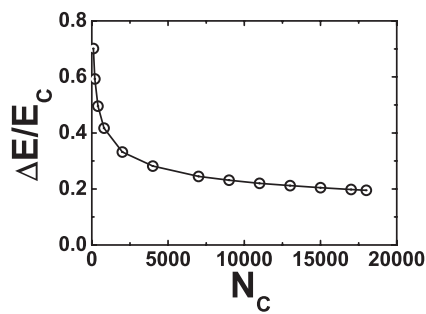}
\caption{Average energy level spacing of non-resonant states near
$E=0$ as a function of $N_c$ at $B=1.9$T ($\alpha=1$). The dimension
$N_c$ varies from $101$ to $18001$. Matrix sizes $N_c$ are $101$,
$201$, $401$, $801$, $2001$, $4001$, $9001$, $11001$, $15001$, and
$18001$.}\label{fig:spacing}
\end{center}
\end{figure}

The  energy spectrum of $J=-1/2$  in Fig.\ref{fig:Energy_spec}(a)
also display non-resonant states. Probability density of a
non-resonant state looks qualitatively different from that of the
corresponding unrenormalized LL state. The probability density of a
non-resonant state $|\Psi_{-1}^{-1/2}(r)|^2$ at $B=2.04$T
($\alpha=0.9$) is shown in Fig.\ref{fig:Anom}(a).  Its  wavefunction
has a {\it large}  peak at $r=0$, which is different from the usual
behavior of the wavefunction a non-resonant state. As $\alpha$
increases the peak at $r=0$ increases even more and the state becomes anomalous. 
Non-resonant  states are  unique to graphene parabolic wells and do
not exist in ordinary parabolic wells. In the absence of the
parabolic potential its energy is $E=-\sqrt{2}E_C=-0.051$eV while in
the presence of the potential it is
$\epsilon_{-1}^{-1/2}(2.04)=1.5E_C=0.054$eV. Fig.\ref{fig:spacing}
displays the energy level spacing of non-resonant states  as a
function of $N_c$ at $\alpha=1$ ($B=1.9$T).  We observe that the
level spacing decreases rather slowly for large $N_c$. However, the
energies of resonant quasi-boundstate converge rather quickly, see
Fig.\ref{fig:collap}.

\subsection{Anomalous states}

\begin{figure}[!hbpt]
\begin{center}
\includegraphics[width=0.3\textwidth]{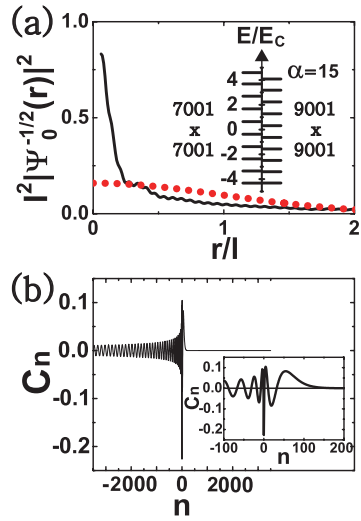}
\caption{(a) Total probability density  of $|\Psi_0^{-1/2}\rangle$
with  energy $0.177$eV at $B=0.312$T ($\alpha=15$) (solid).
Corresponding state at $\alpha=0$ is shown as dotted line. The
eigenstate is obtained by diagonalizing $7001\times 7001$
Hamiltonian matrix. Energy levels are also shown at $B=0.312$T
($\alpha=15$) for $N_c=7001$ and $9001$.   (b) Expansion
coefficients $C_n=\langle n|\Psi_0^{-1/2}\rangle$ of anomalous
eigenstate $|\Psi_0^{-1/2}\rangle$ .} \label{fig:Ano}
\end{center}
\end{figure}

The  energy spectrum of $J=-1/2$  in Fig.\ref{fig:Energy_spec}(a)
also display anomalous states at low magnetic fields of the strong
coupling regime $\alpha\gg 1$.  In this  regime  both resonant and
non-resonant states transform into {\it anomalous} states, and a
sharp distinction between resonant and non-resonant states  no
longer exists.  Such a state is shown in Fig.\ref{fig:Ano}(a) with
the energy $E=0.177$eV (in unit of $E_C$ it is $12.6$). We see in
Fig.\ref{fig:Ano}(a) that the peak value of probability density at
$r=0$ is much larger than the unperturbed value of
$\frac{1}{2\pi}\simeq 0.16$.\cite{com2}. For this state  the
penetration into the barrier should start from the turning point
$r_b$ satisfying $\frac{1}{2}\kappa r_b^2=E$. From this we find that
the value $r_b/\ell$ is $1.3$, which is rather different from the
estimate of about $0.3$ obtained from numerical result shown in
Fig.\ref{fig:Ano}(a) (note that  the probability density under the
barrier oscillates).  An anomalous state is a strong coupling effect
and can only be obtained correctly by computing large Hamiltonian
matrices.

Fig.\ref{fig:Ano}(b) displays the expansion coefficient $C_n$ as a
function of $n$.  Note that $C_n$ has a long oscillating tail for
$n<0$.  This is intimately related to the probability density having
a long oscillating tail under the barrier.  The sum of $C_n$ for
$n<0$ is approximately zero while the sum for $n\geq 0$ is finite
and makes $\Psi_0^{-1/2}(0)$ large (see Eq.(\ref{mhalf})). Note that
the probability density under the barrier  is somewhat smaller than
that of $\alpha=0$.

\begin{figure}[!hbpt]
\begin{center}
\includegraphics[width=0.3\textwidth]{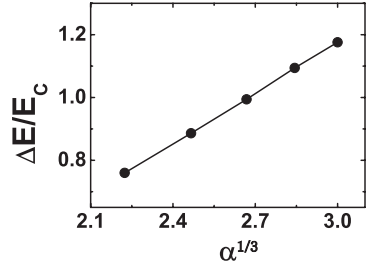}
\caption{Dimensionless average energy level spacing of the energy
spectrum of Hilbert subspace $J=-1/2$ in the strong coupling regime.
We have used $N_c=9001$.} \label{fig:levspacing}
\end{center}
\end{figure}

When $B\rightarrow 0$ or $\alpha\rightarrow \infty$ the natural
length and energy scales of the problem are  $\xi=(\frac{\hbar
v_F}{\kappa})^{1/3}$ and $\kappa^{1/3}(\hbar v_F)^{2/3}$ (Note
$\xi/\ell=\alpha^{-1/3}$). In units of $E_C$ this energy scale is
$\alpha^{1/3}$, which should be proportional to the dimensionless
energy level spacing of the Hilbert subspace of $J$ in the strong
coupling regime. Our numerical results in the strong coupling regime
$\alpha\gg 1$ are indeed consistent with this, see
Fig.\ref{fig:levspacing}.

\section{Scaling of optical transitions}

\subsection{Scaling results}

\begin{figure}[!hbpt]
\begin{center}
\includegraphics[width=0.3\textwidth]{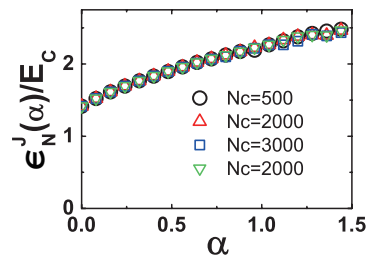}
\caption{Data collapse of dimensionless energies of quasi-boundstate
$\epsilon_{1}^{1/2}(\alpha)$   for different values of $N_c$ and
$\kappa$.  Circles, triangles, and squares are for $\kappa
=0.1$meV/nm$^2$.  Inverted tiangles are for $\kappa
=0.2$meV/nm$^2$.}\label{fig:collap}
\end{center}
\end{figure}

\begin{figure}[!hbpt]
\begin{center}
\includegraphics[width=0.3\textwidth]{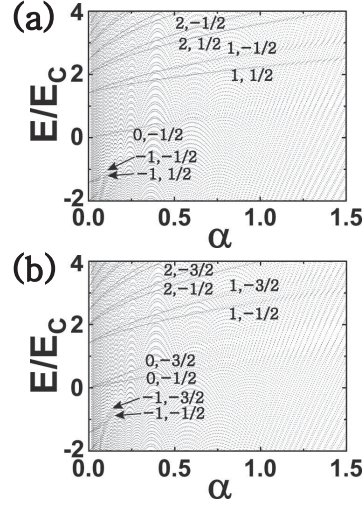}
\caption{(a) Energy spectra for $ J=-\frac{1}{2}$ and $\frac{1}{2}$
plotted together. (b) Energy spectra for $ J=-\frac{1}{2}$ and
-$\frac{3}{2}$ plotted together.  Matrix sizes $N_c$ are  $2001$ for
$ J=-\frac{1}{2}$, $2001$ for $ J=-\frac{3}{2}$, and $2000$ for
$J=\frac{1}{2}$.}\label{fig:opt}
\end{center}
\end{figure}

In the previous section we showed that energies of the resonant
quasi-boundstates  depend on {\it both}  $\kappa$ and $B$. Here we
will show that their energies, when measured in units of $E_C$ in
the limit of large $N_c$, follow a scaling function of a {\it
single} dimensionless variable, namely, the dimensionless coupling
constant $\alpha$, see Fig.\ref{fig:collap}. Fig.\ref{fig:opt}(a)
displays the dimensionless energies of resonant quasibound states
$\epsilon_{N}^{J}(\alpha)$  as a function of $\alpha$ for $J=-1/2$
and $1/2$. Fig.\ref{fig:opt}(b) displays similar results for
$J=-3/2$ and $-1/2$.

Since dimensionless energies of  resonant quasi-boundstates
$\epsilon_{N}^{J}(\alpha)$  satisfy a scaling function the
transition energies
$E=\epsilon_{N'}^{J'}(\alpha)-\epsilon_{N}^{J}(\alpha)$ between them
also obey a scaling
\begin{eqnarray}
\frac{E}{E_C}=f_{N\rightarrow N'}^{J\rightarrow J'}(\alpha).
\end{eqnarray}
This scaling  relation holds as long as quasi-boundstates are well
defined.

\subsection{Optical transition energies and selection rules}

\begin{figure}[!hbpt]
\begin{center}
\includegraphics[width=0.3\textwidth]{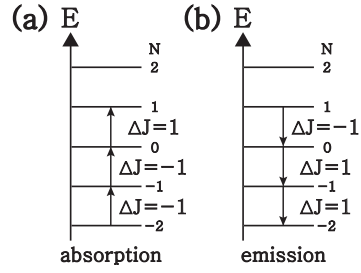}
\caption{We assume that photons are polarized along x-axis. (a) Some
examples of absorption selection rules for LL states in the absence
of a parabolic potential  given by Eq.(\ref{eq:spinor}). In fact
there are also transitions with the selection rule $N\rightarrow
|N|+1$ with $\Delta J=1$ ($N\leq -1$) and $N\rightarrow |N|-1$ with
$\Delta J=-1$ ($N\leq -2$). However, they have high energies and are
not considered here. (b) Some examples of emission selection rules
are shown. Selection rules $N\rightarrow -|N|-1$ with $\Delta J=1$
($N\geq 1$) and $N\rightarrow -|N|+1$ with $\Delta J=-1$ ($N\geq 2$)
are also possible. These transitions are not relevant
here.}\label{fig:trans}
\end{center}
\end{figure}

Before we compute strengths and selection rules of optical
transitions let us first mention some useful results in computing
them. First, in the absence of a parabolic potential absorption
selection rules are $\epsilon_{N}^{-1/2}(\alpha)\rightarrow
\epsilon_{N+1}^{1/2}(\alpha)$ for  $N\geq 0$ and
$\epsilon_N^{1/2}(\alpha)\rightarrow \epsilon_{N+1}^{-1/2}(\alpha)$
for $N <0$.  These selection rules are displayed schematically in
Fig.\ref{fig:trans} (see also TABLE II in Ref.\cite{Park2}). Due to
mixing of different LL states by the parabolic potential the
selection rule $\Delta N=\pm1$ must be {\it relaxed}. Second, the
energies of the $N=0$ LL states are split by the perturbing
parabolic potential and they increase as $J$ decreases:
$\epsilon_{0}^{-1/2}(\alpha)<\epsilon_{0}^{-3/2}(\alpha)<\epsilon_{0}^{-5/2}(\alpha)<\cdots.$
This is also true for other LL states. Third, the optical
transitions with the strongest strength occur between $J=1/2$ and
$-1/2$. Other transitions, for example, transitions between $J=-3/2$
and $-1/2$ are weaker: the transition
$\epsilon_{0}^{-1/2}(0.25)\rightarrow \epsilon_{1}^{1/2}(0.25)$ has
the strength $0.37$ with transition energy $1.44E_C$ while the
transition $\epsilon_{0}^{-3/2}(0.25)\rightarrow
\epsilon_{1}^{-1/2}(0.25)$  has the strength $0.21$ with transition
energy $1.34E_C$ (the involved energy levels are shown in
Fig.\ref{fig:opt}).  These strengths and the corresponding
transition energies decrease as the value $J$ of initial states of
optical transitions decreases.

\begin{figure}[!hbpt]
\begin{center}
\includegraphics[width=0.5\textwidth]{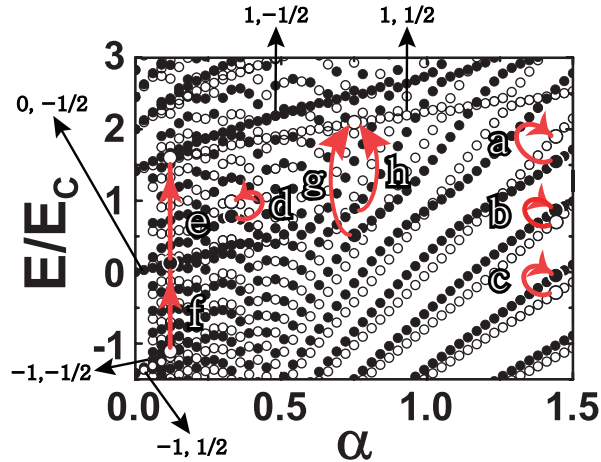}
\caption{ To display clearly possible optical transitions we plot
together, for small $N_c=101$ and $100$, the  energy spectra for
$J=-\frac{1}{2}$ (black dots) and $\frac{1}{2}$(white dots). The
value of the strength of the potential $\kappa=0.1$meV/nm$^2$
corresponds to $\alpha=1.424$ and $\alpha=0.116$ at $B=1.5$T and
$B=8$T, respectively. Quasi-boundstate energies
$\epsilon_N^J(\alpha)$ are labeled by $(N,J)$. }\label{fig:opt2}
\end{center}
\end{figure}

Let us use these results to understand what type of optical
transitions are possible. For this purpose we display in
Fig.\ref{fig:opt2}, for a relatively small value of $N_c$,
eigenenergies for $ J=-\frac{1}{2}$ and $\frac{1}{2}$ (Results are
qualitatively similar to those of a large value of $N_c$).
Absorption transition strength between initial and final states
$\Psi_i$ and $\Psi_f$ is $|\langle\Psi_f|\sigma_x|\Psi_i\rangle|^2$.
Note that there are possible transitions from resonant
quasi-boundstates of negative energies, which is a new feature. An
example of this transition is shown  as (f)  in Fig.\ref{fig:opt2}.
Its strength is $0.386$. Also transitions between resonant
quasi-boundstates of positive energies are possible.  An example  is
shown  as (e) with the strength $0.461$. Energy of a resonant
quasi-boundstate will split into several values at small  $B$ due to
anticrossing with other levels. This will lead to a splitting of
transitions. An example is shown as (g) and (h). Some examples of
absorption transitions involving non-resonant states are also shown.
For these transitions, due to mixing of different LL states by the
parabolic potential, the selection rule $\Delta N=\pm1$ must be
relaxed. Examples of these transitions are shown as (a), (b), (c),
and (d), in Fig.\ref{fig:opt2}. Their absorption strengths are
$0.255$ (a), $0.316$(b), $0.202$ (c), and $0.207$ (d).

The next dominant absorption transitions occur for
$(J,J')=(-3/2,-1/2)$.  In the absence of the parabolic potential the
selection rules are $\epsilon_{N}^{-3/2}(\alpha)\rightarrow
\epsilon_{N+1}^{-1/2}(\alpha)$ for $N\geq 0$ and
$\epsilon_N^{-1/2}(\alpha)\rightarrow \epsilon_{N+1}^{-3/2}(\alpha)$
for $N <0$.   Similar selection rules hold for other possible
$(J,J')$ with $\Delta J=\pm 1$.

\section{Summary}

In each Hilbert subspace of angular momentum $J$ we have studied how
eigenvalues and eigenstates of a parabolic dot in a magnetic field
evolve as $\alpha$ increases.  We have found that they change in a
non-trivial way. In the weak coupling limit of $\alpha\rightarrow 0$
one recovers discrete LL spectrum of graphene. In the intermediate
coupling regime  non-resonant states form a closely spaced energy
spectrum, see Fig.\ref{fig:Energy_spec}(a) (The result is different
from the case of a cylindrical potential, whose the energy spectra
in a magnetic field are discrete without
quasi-boundstates\cite{Park2,Park1,Rec} except at $B=0$\cite{Mat}).
In addition, we find, counter-intuitively, that resonant
quasi-boundstates of both positive and negative energies exist in
the spectrum, see Fig.\ref{fig:Energy_spec}(c). Closely spaced
spectrum is consistent with the presence of resonant
quasi-boundstates\cite{Landau}. The presence of resonant
quasi-boundstates of negative energies is a unique property of
massless Dirac fermions, but they are well-defined  only for $\alpha
< 1$.  In the  strong coupling regime of $\alpha\gg 1$ all resonant
and non-resonant states become anomalous states that develop  a
sharp peak in the well and decay slowly with small oscillations
under the barrier. The average energy level spacing in each Hilbert
subspace of $J$ approaches a constant value. However, note that the
total density of states is the sum of each density of states
computed in different Hilbert subspaces. When $\alpha$ is too large
the coupling between $K$ and $K'$ valleys may have to be
included\cite{JWLee}.

Optical transitions from resonant quasi-boundstates of positive
energies are possible.  There are also possible transitions from
resonant quasi-boundstates of negative energies, which is a new
feature.  Moreover, we  find that transition energies between
resonant quasi-boundstates follow a scaling as a function of the
coupling constant $\alpha$. Absorption transitions involving
non-resonant states are also possible, and for these transitions the
selection rule $\Delta N=\pm1$ must be relaxed due to mixing of
different LL states by the parabolic potential.   It would be also
interesting to observe experimentally the splitting of optical
transition energies due to anticrossing of resonant
quasi-boundstates with other states.

\ack This research was supported by Basic Science Research Program
through the National Research Foundation of Korea (NRF) funded by
the Ministry of Education, Science and Technology
(2012R1A1A2001554). We thank H. W. Lee for several useful
suggestions.


\begin{thebibliography}{10}
\bibitem{com1} In the strong coupling limit  many levels with
different values of $n$ are heavily mixed.
\bibitem{com2}  For sufficiently large $\alpha$ even eigenstates with
eigenvalues close to zero have narrow probability densities near
$r=0$.
\end{thebibliography}
\end{document}